# Jigsaw model of the origin of life

John F. McGowan III[*]

GFT Group, 1379 Snow Street, Suite 3, Mountain View, CA 94041

## ABSTRACT

It is suggested that life originated in a three-step process referred to as the jigsaw model. RNA, proteins, or similar organic molecules polymerized in a dehydrated carbon-rich environment, on surfaces in a carbon-rich environment, or in another environment where polymerization occurs. These polymers subsequently entered an aqueous environment where they folded into compact structures. It is argued that the folding of randomly generated polymers such as RNA or proteins in water tends to partition the folded polymer into domains with hydrophobic cores and matching shapes to minimize energy. In the aqueous environment, hydrolysis or other reactions fragmented the compact structures into two or more matching molecules, occasionally producing simple living systems, also known as autocatalytic sets of molecules. It is argued that the hydrolysis of folded polymers such as RNA or proteins is not random. The hydrophobic cores of the domains are rarely bisected due to the energy requirements in water. Hydrolysis preferentially fragments the folded polymers into pieces with complementary structures and chemical affinities. Thus the probability of producing a system of matched, interacting molecules in prebiotic chemistry is much higher than usually estimated. Environments where this process may occur are identified. For example, the jigsaw model suggests life may have originated at a seep of carbonaceous fluids beneath the ocean. The polymerization occurred beneath the sea floor. The folding and fragmentation occurred in the ocean. The implications of this hypothesis for seeking life or prebiotic chemistry in the Solar System are explored.

**Keywords:** RNA, protein, polymer, jigsaw, origin

## 1. INTRODUCTION

The simplest living organisms on present-day Earth are complex systems of interacting, often tightly coadapted molecules. No hypothetical self-replicating molecule such as a self-replicating RNA molecule, a current favorite candidate for the first life form, has been found in nature or synthesized in the laboratory. DNA, for example, replicates only with the aid of DNA polymerase molecules and other supporting machinery found in cells. Viruses are not capable of independent replication and are usually made of additional molecules besides RNA or DNA.

The most popular models of the origin of life propose some form of a prebiotic soup[1,2,3,4,5,6,7,8]. The prebiotic soup refers to a primordial ocean filled with simple organic molecules such as amino acids. Polymers such as proteins break down in water rather than polymerize. Polymerization requires dehydration, a two-dimensional surface, or special molecules that catalyze polymerization. Prebiotic soup models often propose that the building blocks polymerized to form a first self-replicating molecule that evolved into more complex life. Many other models of the origin of life have been proposed. It has been suggested that life evolved from clay replicators[9]. It has been suggested that life on Earth originated from another world or from deep space[10,11]. It has been proposed that life originated in primordial hydrocarbons such as oil and natural gas trapped in the Earth's crust[12]. In general, there are many difficulties with origin of life models as currently conceived[13].

One possible solution to the absence of self-replicating molecules and the complexity of even the simplest living organisms is that the first living organism was a system of interacting molecules, an autocatalytic set of molecules. It has been proposed, for example, that autocatalytic sets of molecules would have formed inevitably in a prebiotic soup[14]. In this model, species of molecules were produced by random polymerization in the hypothetical prebiotic soup. If a sufficient number of different species of molecules are present in the prebiotic soup, an autocatalytic reaction is always possible. This autocatalytic reaction makes copies of its constituent molecules and consumes the prebiotic soup.

---

* Correspondence: E-Mail: jmcgowan@veriomail.com; Telephone: (650) 961-1429

Biological molecules such as DNA, RNA, and proteins have complementary shapes and chemical affinities that appear to be necessary for the proper functioning of living organisms. It is not clear that independent, random synthesis of organic molecules in a hypothetical prebiotic soup could generate an autocatalytic set of molecules. The tighter the match required for the autocatalytic set to function, the less probable the synthesis of an autocatalytic set by independent random polymerization of amino acids or other monomers. The explanation for the origin of life may lie in chemical or physical processes that preferentially or exclusively synthesize systems of tightly matched, interacting molecules.

## 2. METHODS TO MAKE SYSTEMS OF INTERACTING MOLECULES

### 2.1 Molecular templates

In mechanical systems, a number of mechanisms are known that make systems of matching mechanical parts. Human beings use all of these mechanisms to design and build machines. At least some occur in nature in the absence of intelligence. The simplest mechanism is to use one constituent part of a machine as a template for a second matching part. Typically, a matching part is formed by a fluid or gel that conforms to one or more side of the first part. The fluid hardens into a solid, forming a matching part. Polymers such as latex, polyurethane, and silicone are commonly used to make molds and matching parts by this method. Metal casting involves a similar process. Lava provides a natural example of this process. A volcano with a lava plug sealing the crater is a naturally occurring machine, a cannon, made of two tightly matched parts. The lava plug conforms to the surface of the volcano crater, trapping gases below. Without the tight match, the gases may be released slowly and non-destructively. The tight match of the shapes causes an accumulation of pressure and an explosion, projecting the lava plug into the sky, perhaps into orbit or deep space. Mud, clay, and water freezing into ice can make matching parts through a template process. The folding of an organic polymer such as a protein guided by an already folded polymer, a "chaperone" molecule, may be a biochemical template process.

In some cases, the surface of a folded organic polymer may act as a template that catalyzes the polymerization of a matching surface, a polymer in a folded configuration. In the case of standard proteins, the matching surface would be made of $\alpha$ helices, $\beta$ sheets, open loops, or other secondary structures. In the case of proteins, these matching surfaces would probably lack hydrophobic cores making the surfaces unstable. The problem for the origin of life is that, in general, at least two molecular templates are needed to make a stable folded three-dimensional molecule. The living system must make the molecular templates as well. Some additional mechanism is probably needed.

### 2.2 Molecular building blocks and L-systems

Another mechanism to make systems of mechanical parts is the assembly of mechanical systems from standardized simpler matching parts. Extremely complex systems can be constructed from a relatively small number of basic building blocks. The popular Lego toy is an example of this mechanism. Builders and engineers use the same method. It is not clear that this mechanism occurs in nature in the absence of intelligence. However, mobile genetic elements that carry a genetic payload and that copy or insert themselves or the payload alone in place of or next to a specific target DNA sequence can implement this mechanism within living organisms. The genetic payload must consist of modular building blocks of genes such as the exons found in eukaryotes, regulatory binding sites, or the DNA sequences targeted by the mobile genetic elements.

Targeted mobile genetic elements that carry genetic payloads can implement Lindenmayer-systems or L-systems, a mathematical model of plant growth and cell differentiation proposed by the late biologist Aristid Lindenmayer[15]. The central concept of L-systems is that of rewriting. In general, rewriting is a technique for defining complex objects by successively replacing parts of a simple initial object using a set of rewriting rules or productions[16]. An L-system is a collection of rewriting rules that replace a symbol or group of symbols with a different, usually more complex set of symbols.

A simple example illustrates how a system of targeted mobile genetic elements with genetic payloads can build complex networks of genes. The system consists of a target DNA sequence identified by the symbol $\alpha$ for a network or sub-network of interacting genes, a target DNA sequence identified by the symbol *I* for interface, an exon *P* coding for a folding domain P, a regulatory binding site *Q* recognized by the folding domain P, an exon *R* coding for a folding domain R, and a regulatory binding site *S* recognized by the domain coded by *R*. A semicolon represents a stop codon separating two genes. Each mobile genetic element implements a single rewriting rule in the L-system. The domains implement a linear interface system similar to a lock and key. For example, the protein PPR, consisting of the domains P, P, and R in a linear chain, binds only to the matching regulatory binding sites QQS (Figure 1). This linear modular system may be similar to the regulatory proteins made of modular zinc finger motifs arranged in tandem and their associated regulatory binding sites[17].

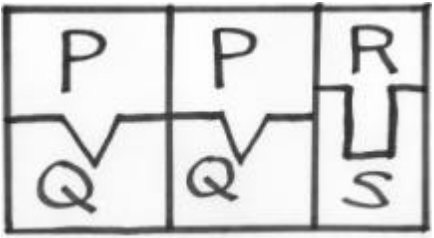

**Figure 1** Modular Three Domain Proteins PPR and Regulatory Binding Sites QQS Fit Together

The targeted mobile genetic elements with genetic payloads can be represented symbolically as rewriting rules in an L-system. This is a stochastic L-system in which multiple rewriting rules have the same left hand side − *I* and α in the example. Which rule is used is random or stochastic. Amongst other advantages, this permits different target sequences *I* to represent distinct and unique interfaces in the network.

$$a \circledR d \qquad\qquad\qquad I \circledR RIS$$

$$a \circledR I \qquad\qquad\qquad I \circledR P;Q$$

$$I \circledR PIQ \qquad\qquad\qquad I \circledR R;S$$

Within the genome, a targeted mobile genetic element with a genetic payload might look something like:

*(DNA sequence for a transposase protein that cuts or copies DNA delimited by the transposition start and stop codes and splices at the target DNA sequence such as I) (Transposition start code) (genetic payload of exons, regulatory binding sites and DNA target sequences such as PIQ)(transposition stop code)*

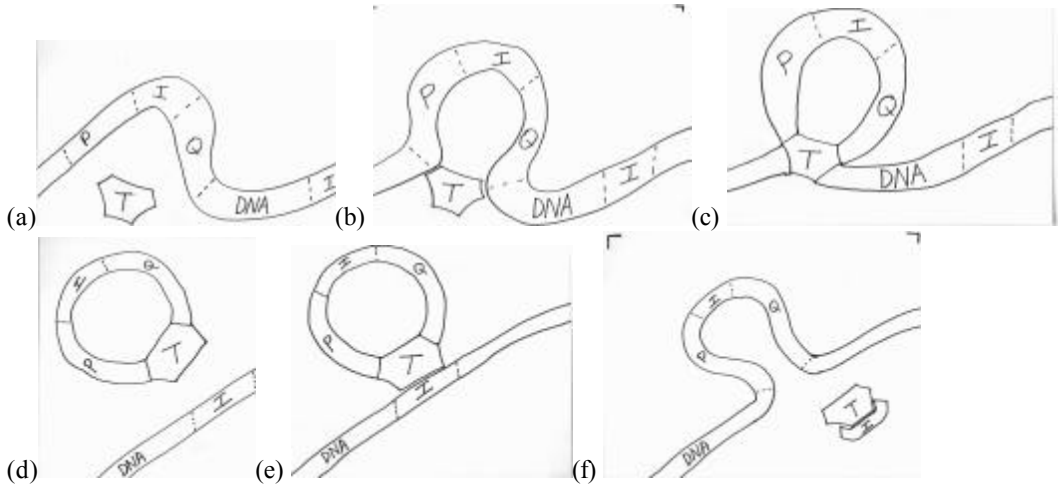

**Figure 2** A Transposase T Cuts DNA Segment *PIQ* and Splices at *I* (*I → PIQ*)

Figure 2 illustrates a transposase T, a special protein that cuts and splices a DNA segment, cutting the segment *PIQ* and splicing the segment in place of the target DNA sequence *I*. When any of the mobile genetic elements that targets the DNA sequence *I* in the genome encounters the target DNA sequence *I*, it will substitute its genetic payload, represented by the right hand side of the rewriting rule, into the genome. The process continues until the last target DNA sequence *I* is consumed. The recursive nature of the rewriting rule, the mobile genetic element, where *I* appears on both the left and right hand side of the rule enables the rule to build arbitrarily complex systems.

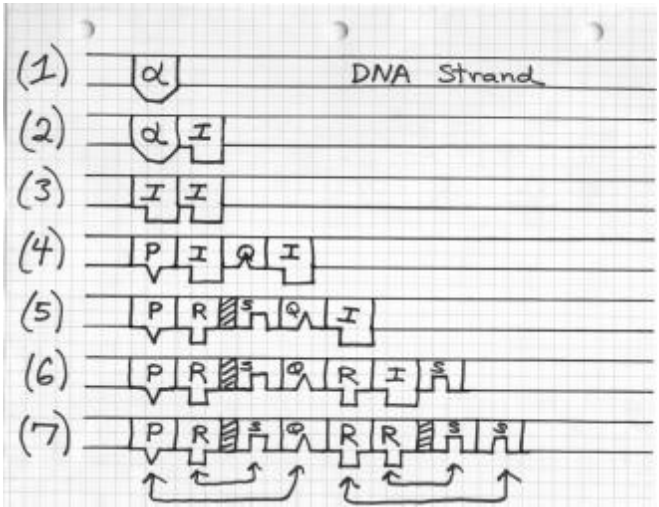

| Step | Rule |
|------|------|
| (1) → (2) | $\alpha \rightarrow \alpha I$ |
| (2) → (3) | $\alpha \rightarrow I$ |
| (3) → (4) | *I ⑩PIQ* |
| (4) → (5) | *I ⑩R;S* |
| (5) → (6) | *I ⑩RIS* |
| (6) → (7) | *I ⑩R;S* |

**Figure 3** Repeated Substitutions by Targeted Mobile Genetic Elements Build a Network of Genes

Figure 3 illustrates the mobile genetic elements expanding an initial target DNA sequence α into a network of three genes. In step one, the mobile genetic element that implements α → αI replaces α with αI. In step two, the mobile genetic element that implements $\alpha \rightarrow I$ replaces αI with II in the DNA strand. The process continues until the network of three genes is constructed. This simple example molecular L-system can build larger networks of hundreds or thousands of genes. More sophisticated molecular L-systems can build networks with branches, feedback loops, and other complex structures.

Repeated insertions or substitutions of segments of genes such as the exons and regulatory binding sites by targeted mobile genetic elements with genetic payloads can build complex networks of structural proteins, regulatory proteins, and regulatory binding sites such as the signal transduction networks that include the oncogenes and tumor suppressor genes that are often mutated in cancer[18]. A mutation or activation of these hypothetical mobile genetic elements would restructure the network extensively, changing many genes simultaneously in a coordinated fashion and probably causing substantial chromosomal rearrangements as the mobile genetic elements cut and splice large sections of DNA. This provides a possible mechanism for saltatory evolution[19]. While the view that this actually happens in living organisms is not generally accepted, it is suggested in the work of the late Barbara McClintock, James Shapiro, and some other researchers into mobile genetic elements[20,21,22,23,24,25,26].

Molecular L-systems need simple molecular building blocks with complementary shapes and chemical affinities − like P, Q, R, S, α, I, and the transposases that implement the rewriting rules in the example − to get started. To explain the origin of life with molecular L-systems a mechanism to make the building blocks with complementary shapes and chemical affinities is needed.

## 2.3 Molecular jigsaw mechanism

A third mechanism to make systems of matching mechanical parts is to fragment or cut a larger parent piece into two or more smaller, exactly matching pieces. This mechanism is used routinely to design and make jigsaw puzzles made of hundreds or thousands of pieces. A natural example is the fragmentation of rock into the system of fissures and cavities comprising a geyser by heat and water upwelling from deep in the Earth. Geysers are natural machines in which the constituent parts must be tightly matched to yield the intermittent and sometimes periodic eruptions of hot water and steam. Geysers are extremely rare with only about two hundred known. Three geysers are periodic "Old Faithful" geysers[27,28,29]. The fragmentation of a folded organic polymer such as a protein by hydrolysis or other chemical reactions may be a molecular jigsaw mechanism. A molecular jigsaw mechanism is the best candidate for an explanation of the origin of life of the methods suggested. Alone of the suggested methods, it can make a system of matching, interacting molecules in a single, simple step.

## 3. THE JIGSAW MODEL

### 3.1 Basic concept

The jigsaw model of the origin of life assumes that random polymerization of organic monomers such as amino acids occurred as in many prebiotic soup models. These polymers then entered an aqueous environment where those polymers with protein-like properties folded into three-dimensional structures. For simplicity, the following discussion will assume the polymers were standard proteins. When a long protein chain is folding, two major forces are competing. The protein attempts to fold to bury the hydrophobic amino acid residues in the interior, isolated from water. This forms folding domains with hydrophobic cores that may be relatively stable. At the same time, the folding domains must interact with their geometrical neighbors. The minimum energy configuration may be achieved locally by domains with complementary shapes and chemical affinities. The folding of a protein may therefore be a partitioning process where amino acid residues are partitioned among folding domains to achieve both hydrophobic cores and complementary shapes and chemical affinities.

Hydrolysis of a folded protein avoids bisecting the hydrophobic cores of the folded domains, which requires much energy. Instead, hydrolysis tends to attack the open loops that tie separate domains together. It seems likely that hydrolysis of a folded protein often fragments the protein into the folded domains, yielding a system of matched, interacting molecules, with complementary shapes and chemical affinities.

In general, a system of interacting proteins will not be a living system, an autocatalytic set. There is at least one scenario in which the proteins produced by hydrolysis as suggested above may be alive. A common simple folding domain in proteins is a four-helix bundle made of four tightly packed α helices. An α helix is a common secondary structure in proteins in which the backbone of the protein is wound in a helix and the amino acid residues point outward. In a four-helix bundle, the hydrophobic residues of the four α helices are buried in the interior of the four-helix bundle and form a hydrophobic core. The parent protein is made of several four-helix bundles that are tightly packed together with tightly matching shapes and chemical affinities. When hydrolysis fragments this parent protein, the several four-helix bundles separate. If each four-helix bundle can catalyze the polymerization of the matching faces, two α helices, of its neighbors, the system can become an autocatalytic set of proteins.

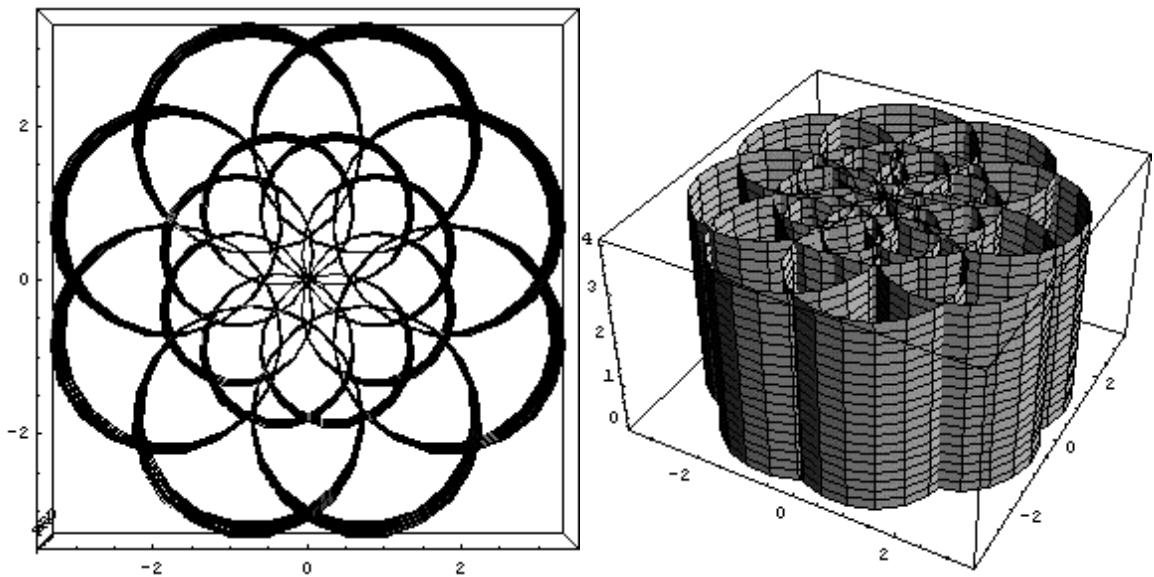

**Figure 4** Parent Protein Made of Tightly Packed α Helices (Top and Perspective Views)

Figure 4 illustrates a parent protein made of tightly packed α helices. The protein is a single long chain folded into a complex structure. Each cylinder represents an α helix. The α helices are linked by short open loops at the ends (not shown). The α helices overlap. The ridges of one α helix fit precisely into the grooves of the adjacent α helix. The α helices form four wedge-shaped four-helix bundles W, X, Y, and Z. Each four-helix bundle has a hydrophobic core. The adjacent faces of the four-helix bundles fit together. The larger diameter outer α helices are made of amino acids with large, bulky residues. The smaller diameter inner α helices are made of amino acids with small, compact residues.

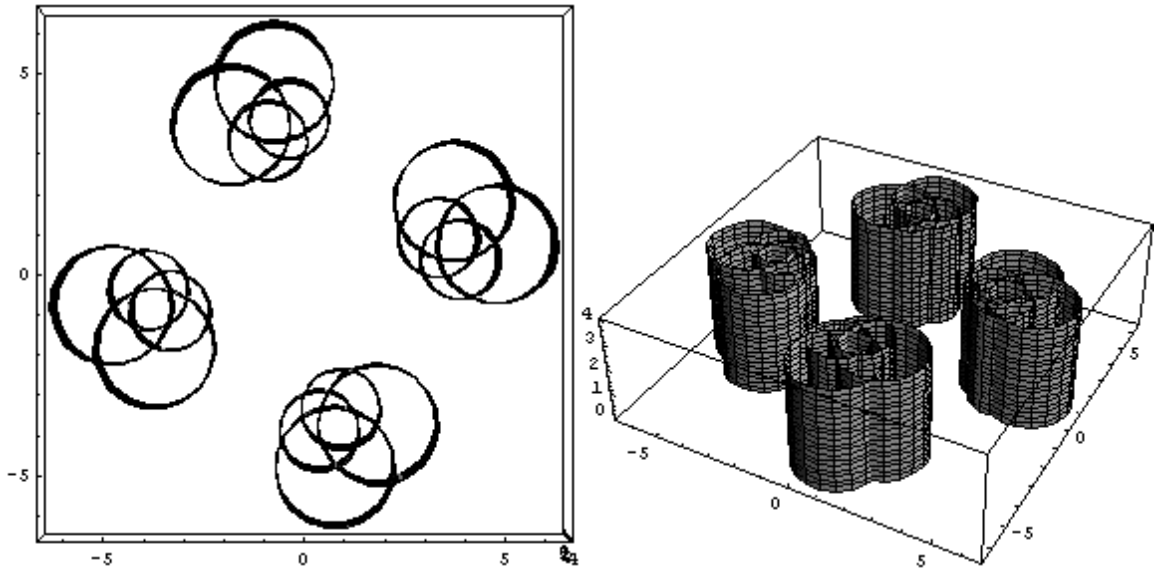

**Figure 5** Hydrolysis Fragments the Parent Protein into Four Wedge-Shaped Four-Helix Bundles

It is difficult for hydrolysis to bisect the hydrophobic cores of the four-helix bundles. In Figure 5, hydrolysis cuts the open loops linking the non-hydrophobic faces of the four-helix bundles. A single open loop connects each four-helix bundle to its neighbor. The parent protein fragments into four separate wedge-shaped four-helix bundles W, X, Y, and Z.

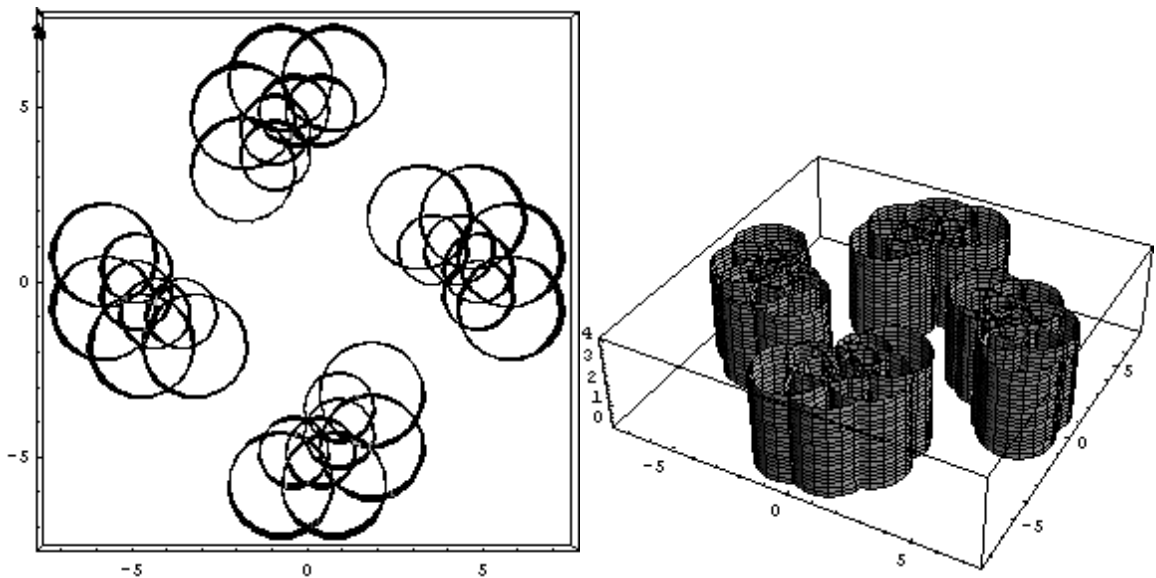

**Figure 6** Wedge-Shaped Four-Helix Bundles Catalyze the Polymerization of the Matching α Helices

Figure 6 illustrates the four-helix bundles catalyzing the polymerization of the matching α helices that form the matching faces of the adjacent four-helix bundles. The four-helix bundles have holes, indentations, and grooves that accept the specific amino acids that comprise the matching α helices. The four-helix bundle catalyzes the polymerization of the matching face of the adjacent four-helix bundle. The wedge-shaped four-helix bundles turn into one half of the original parent protein. The newly polymerized α helices have exposed hydrophobic residues. The matching halves will tend to stick together to bury the hydrophobic residues (Figure 7).

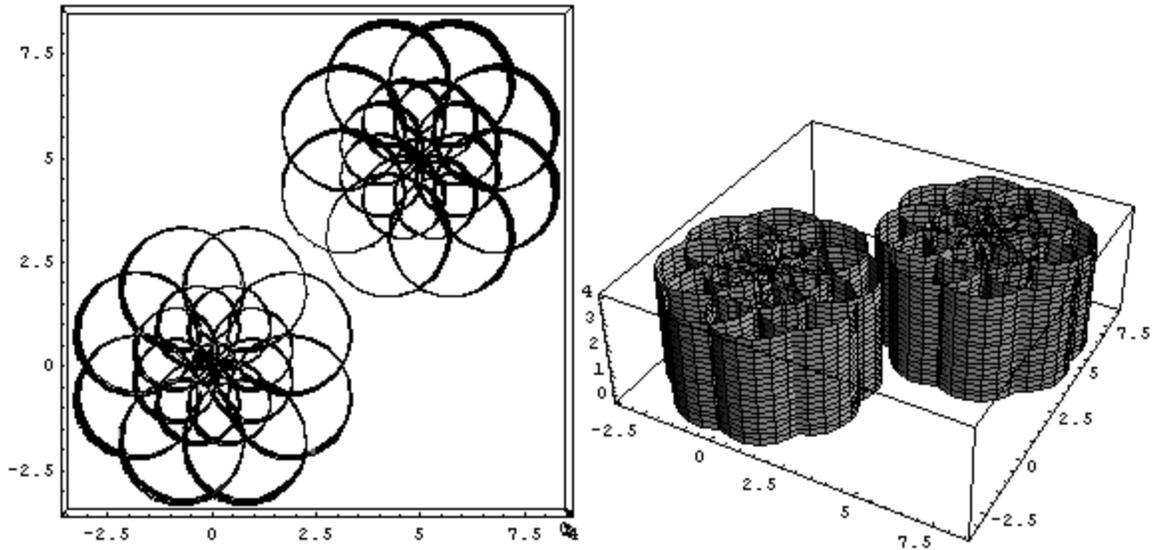

**Figure 7** The Halves Produced by Polymerization Collide and Combine into New Proteins.

For illustrative purposes, the open loops that connect the ends of the α helices in the four-helix bundles have been neglected so far. The example four-element (each element is a four-helix bundle) autocatalytic set can catalyze formation of these loops if the north end of each element fits into the south end. North and south are defined as the two ends of the major axis of the α helix. Hydrolysis of a folded parent protein may make four-helix bundles with matching north and south ends if the bundles wrap around in a loop in the folded parent protein. More realistically, the autocatalytic set probably has more than four elements. The four elements {W, X, Y, Z} form a disk-shaped segment of a torus (donut shape) shaped three-dimensional region within the parent molecule. Hydrolysis fragments the torus region into an autocatalytic set. Fragments outside the toroidal region are discarded and do not become part of the autocatalytic set.

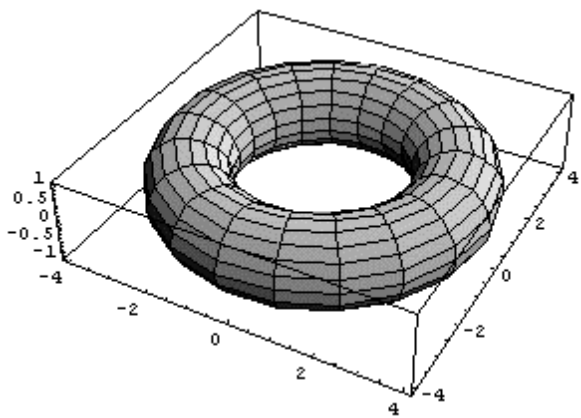

**Figure 8** Torus Made of Autocatalytic Segments (Each Segment is Four Wedge-Shaped Four-Helix Bundles)

Figure 8 illustrates the toroidal parent region of an autocatalytic set made of wedge-shaped four-helix bundles. The torus has twenty disk-shaped segments. Each disk-shaped segment is made of four wedge-shaped four-helix bundles as illustrated in Figures 4,5,6, and 7. Each disk-shaped segment of the torus can catalyze the polymerization of the open loops connecting the ends of the α helices of the adjoining disk-shaped segments.

The example illustrates certain basic principles. The parent protein fragments into simple folding domains with hydrophobic cores such as the wedge-shaped four-helix bundles in the example. The simple folding domains consists of matching faces made of $\alpha$ helices, $\beta$ sheets, or other secondary structures. No secondary structures are buried in the interior of the folding domains. The faces of the folding domain can catalyze the polymerization of matching faces of adjacent folding domains in the original parent molecule. New folding domains are assembled from the new faces. The new faces stick together to bury the exposed hydrophobic residues. The folding domains that make the autocatalytic set form loops, toroids, or similar closed regions within the parent protein. The surface of the region may provide the interface between the autocatalytic set and the environment.

Many simple self-replicating systems such as fire and crystallization in solution are not generally considered alive. All generally accepted living organisms incorporate the functionality of a Turing machine, an abstract model of a programmable computer, through the standard genetic system. Ideally the autocatalytic set should have a genetic system. The current system for encoding proteins in DNA, transcribing the proteins into RNA, and synthesizing the protein from messenger RNA is quite complex. Although hydrolysis of a folded protein may have produced this genetic system during the genesis event, the origin of life, it seems more likely that a simpler genetic mechanism was involved. DNA binding regulatory proteins, regulatory binding sites, and the mobile genetic elements may be descendants of this original system based on simple matching geometry rather than a complex decoder and genetic code based on nucleotide base pairs. The fragmentation of a folded polymer by hydrolysis may have made an autocatalytic set of molecules consisting of the matching building blocks for a molecular L-system including the transposase-like molecules that implement the rewriting rules of the L-system.

### 3.2 The properties of autocatalytic sets

Unlike bacteria and viruses, water-soluble autocatalytic sets of molecules in solution will exhibit a concentration dependent doubling time. If the concentration of the autocatalytic set is below a threshold concentration, hydrolysis will deplete the autocatalytic set faster than it can replenish itself. The autocatalytic set will die. At near-threshold concentrations, the doubling time of the autocatalytic set may be quite long, even years. When the concentration of the autocatalytic set is doubled, the doubling time will be halved or more depending on the number of neighbors needed to catalyze synthesis of a molecule in the autocatalytic set. The differential equations for the concentrations $W(t)$, $X(t)$, $Y(t)$, and $Z(t)$ of the constituents of the autocatalytic set of four proteins {W,X,Y,Z} in the simple example (Figures 4–7) are:

$$dW(t)/dt = kZ(t)X(t) - bW(t) \qquad (1)$$

$$dX(t)/dt = kW(t)Y(t) - bX(t) \qquad (2)$$

$$dY(t)/dt = kX(t)Z(t) - bY(t) \qquad (3)$$

$$dZ(t)/dt = kY(t)W(t) - bZ(t) \qquad (4)$$

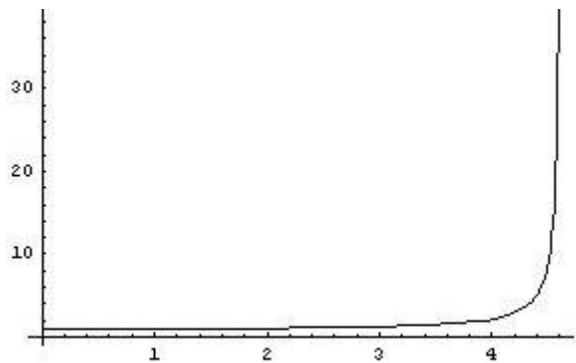

**Figure 9** Concentration of an Autocatalytic Set Versus Time

Figure 9 shows a numerical solution for the concentration $W(t)$ of constituent W of the autocatalytic set computed using the NDSolve function of the Mathematica 4.1 for Microsoft Windows program[30]. The vertical axis is in units of the threshold concentration. The initial concentrations are $W(0) = 1.01$, $X(0) =$ 1.01, $Y(0) = 1.01$, and $Z(0) = 1.01$. The time units are of order the half-life of the protein W in the autocatalytic set. The rate constants $b$ and $k$ are set to one. The rate constant $b$ is the dissociation rate constant due to hydrolysis of the constituents. The constituents W and Y jointly catalyze the polymerization of X and Z as explained above. X and Z catalyze the polymerization of W and Y.

If a water-soluble autocatalytic set of molecules infects a living host, the set may exhibit a long, variable latent period before the onset of detectable symptoms. While the concentration of the autocatalytic set is low in the organism, the organism can easily repair any damage caused by the set digesting tissues in the host. During this period there will be few if any symptoms. This period may last for a long time since the doubling time is long when the concentration is low. However, the set will exponentially increase its concentration as the doubling time is reduced. Then the regenerative powers of the host will be exceeded. Detectable symptoms will appear. Autocatalytic sets may cause slow diseases with variable latent periods followed by the appearance of symptoms. The length of the latent period will depend on the initial dose and concentration of the autocatalytic set in the body.

Because water-soluble autocatalytic sets of molecules can be killed by dilution, it may be difficult to demonstrate that diseases caused by autocatalytic sets are transmissible, infectious diseases. For example, if blood from an animal or human with a disease caused by an autocatalytic set is injected into a healthy animal or human, the autocatalytic set may be diluted and killed in the blood of the healthy animal or human. Casual contact, sweat, saliva, blood, and other fluids may be incapable of transmitting the disease in most cases. The diseases would be acquired by prolonged or repeated exposure to a high concentration of the autocatalytic set such as swimming in a pond containing an active autocatalytic set at high concentration.

## 4. TESTING THE MODEL

### 4.1 Apparatus to detect and isolate water-soluble autocatalytic sets of molecules

The most direct way to test the jigsaw model is to detect and isolate autocatalytic sets of polymers, probably proteins, from the field or a laboratory experiment. An experimental apparatus to detect and isolate water-soluble autocatalytic sets of molecules in samples is possible (Figure 10).

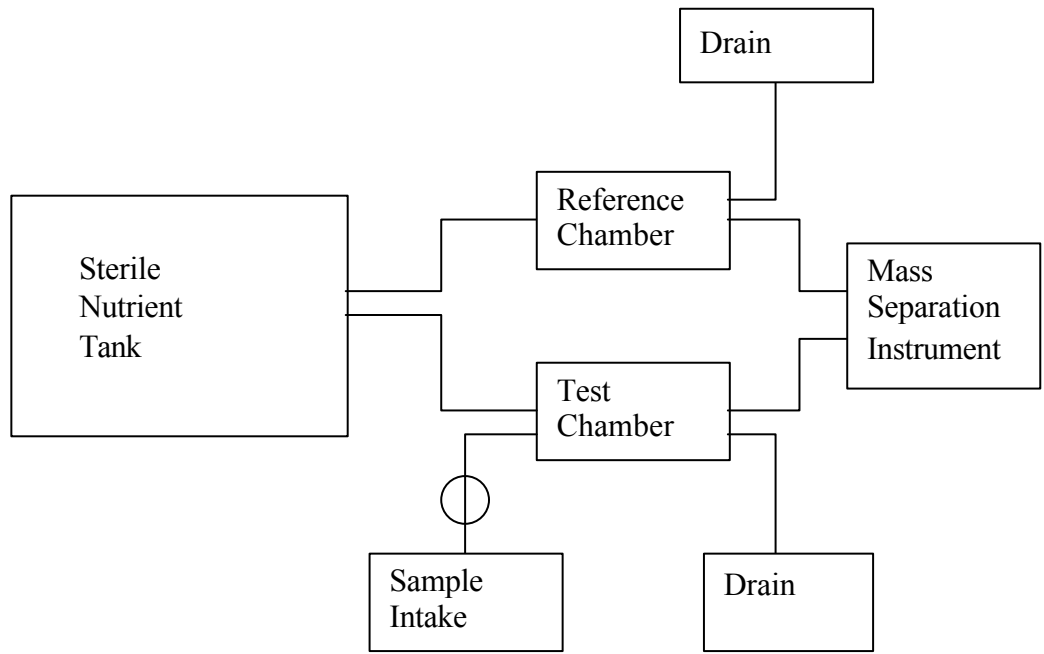

**Figure 10** Apparatus to Detect and Isolate an Autocatalytic Set of Molecules

The test chamber is seeded with a sample from the field or laboratory. The sample is filtered to remove cellular life. If the concentration of the autocatalytic set drops below its critical threshold, the autocatalytic set will die. At low concentrations, the doubling time of the autocatalytic set may be very long, even years, making practical detection of the autocatalytic set impossible. Thus, it is important to start with a sample that probably contains a high concentration of the autocatalytic set and to maintain the high concentration during the detection experiment. The initial high concentration of the autocatalytic set needed for rapid doubling of the set may be achieved by partial evaporation of the test sample, probably using low air pressure to avoid heating that may damage or denature the set.

The test and reference chambers are monitored with the mass separation instrument. The mass separation instrument may be a mass spectrometer (MS), Gas Chromatograph Mass Spectrometer (GCMS), High Performance Liquid Chromatograph Mass Spectrometer (HPLC-MS), gel electrophoresis instrument, other instrument, or some combination of these instruments. Mass spectra are taken periodically, such as once per hour, during the experiment. The chamber is slowly drained of the sample and slowly refilled with the sterile nutrient. The slow removal of the sample and the slow replacement with the sterile nutrient is essential to maintain a high concentration of the autocatalytic set and a short doubling time. The time to replace the sample volume must be longer than the doubling time of the autocatalytic set. Otherwise, the concentration of the autocatalytic set will drop and the set will eventually die.

A mass spectrum due to an autocatalytic set of molecules will persist. A reactant or a catalyst in the original sample from the field or laboratory can cause a characteristic mass spectrum. If a reactant is present, the reactant will be consumed. The characteristic mass spectrum will be diluted away to nothing by the slow flushing and refilling of the test chamber with sterile nutrient. The characteristic mass spectrum will not appear in the reference chamber, which is never exposed to the sample. A catalyst will be diluted by the experiment. The mass spectrum will vanish as the catalyst is diluted. Eventually the catalyst will be diluted to essentially zero and the characteristic mass spectrum will no longer appear in the test chamber. An autocatalytic set of molecules can thus be distinguished from either reactants or catalysts in the original sample. The test and reference chambers should be monitored with microscopes, infrared spectrometers, and other passive instruments. Microscopes can detect contamination by bacteria or other organisms that can cause a false positive signal. In time, the test chamber will contain only the isolated, purified autocatalytic set of molecules and the sterile nutrient solution.

This apparatus can detect and isolate water-soluble autocatalytic sets of molecules and living organisms too small to be detected with optical or even electron microscopes. The next and final section discusses promising places to seek autocatalytic sets of molecules.

## 4.2 Places to seek autocatalytic sets of molecules

Several similarities exist between the expected properties of water-soluble autocatalytic sets of proteins and the reported properties of the infectious agents for transmissible spongiform encephalopathy (TSE) diseases such as kuru, Creutzfeldt-Jakob disease (CJD), Gerstmann-Straussler-Schinker (GSS) disease, scrapie, transmissible mink encephalopathy, chronic wasting disease of mule deer and elk, feline spongiform encephalopathy, and bovine spongiform encephalopathy (BSE, also known as "mad cow disease")[31,32,33,34,35,36]. These diseases have long variable latent periods. In some cases, such as scrapie, the length of the asymptomatic latent period of the disease is inversely related to the inferred initial concentration of the infectious agent. The infectious agent for scrapie is apparently immune to agents that destroy nucleic acids such as nucleases, ultraviolet radiation, zinc ions, psoralen photoadducts, and hydroxylamine. Conversely, agents that destroy or modify proteins such as proteases reduce the infectivity of the infectious agent for scrapie. No bacteria or virus has been successfully isolated from these diseases.

The transmissible spongiform encephalopathy diseases are currently attributed to prions. A prion is not a self-replicating protein molecule or an autocatalytic set of molecules. The prion protein PrP is coded by a gene in both healthy and sick animals and humans. The prion protein PrP is hypothesized to have two or more conformations, a pathogenic conformation and a healthy conformation. The pathogenic conformation of PrP converts the healthy conformation of PrP to the pathogenic conformation. Thus the current interpretation of the experimental results from these diseases rules out autocatalytic sets of molecules as the pathogen. Although a Nobel Prize in Medicine was awarded for the prion theory in 1997, the theory is not accepted by all researchers into the transmissible spongiform encephalopathy diseases[37].

In addition to the transmissible spongiform encephalopathy diseases, it may be worthwhile to apply these experiments to cell-free fluids from patients or animals with poorly understood diseases such as multiple sclerosis, systemic lupus erythematosus, rheumatoid arthritis, insulin dependent diabetes mellitus, and Grave's disease. Most of these diseases are attributed to autoimmune processes[38]. An autocatalytic set could cause immune abnormalities that have been interpreted as evidence of an autoimmune reaction. In particular, the immune system may attack cells in the body infected with the autocatalytic set. No viruses or bacteria could be detected with optical or electron microscopy. In principle, viral isolation experiments should have yielded evidence of the autocatalytic sets of molecules in the form of anomalous bands (proteins or protein-like polymers) in electrophoresis gel experiments. The inability to find viral particles through electron microscopy may have resulted in the neglect or misinterpretation of these results. It is also possible that the bands from the autocatalytic set overlap bands from common cellular proteins, rendering the set invisible unless a life-detection experiment is performed.

Another place to look for possible autocatalytic sets is the oils, tars, or tholins produced in various Miller-Urey type origin of life experiments. The jigsaw model suggests that life originates in a complex polymerized mass such as these materials. Carbonaceous chondrites and oil fields have similar chemistry to the Miller-Urey experiments. In particular, marine seeps of oil and gas or oil fields partially filled with water may be appropriate places to look for autocatalytic sets.

The Viking Lander labeled release experiments yielded positive signals. These have generally been interpreted as caused by inorganic oxidants such as hydrogen peroxide, although this interpretation has been challenged[39]. Autocatalytic sets of proteins or protein-like polymers probably can survive long periods of dehydration, unlike terrestrial life, and thus may be able to survive or even thrive in the Martian dust provided liquid water is occasionally present. Proteins can be resistant to ultraviolet light unlike nucleic acids, which are destroyed by UV radiation. As mentioned, the inferred infectious agents of the transmissible spongiform encephalopathy diseases on Earth are resistant to extreme conditions. Thus, Mars is a worthwhile location for these experiments.

## 5. CONCLUSION

Autocatalytic sets of proteins or protein-like polymers may be feasible. Hydrolysis of folded proteins or protein-like polymers may provide a simple mechanism that occasionally makes autocatalytic sets of proteins or other polymers. While synthesis of an autocatalytic set is probably rare, it is much more likely than the astronomical odds often calculated in origin of life research. Although difficult, it is possible to detect and isolate autocatalytic sets of proteins or protein-like polymers in the laboratory or the field.

## ACKNOWLEDGEMENTS

This work was supported in part by GFT Group. Some preliminary work related to this paper was supported in part by the Applied Information Technology Division, NASA Ames Research Center and the NASA Research and Education Network (NREN). This work benefited from discussions with Ray Cowan, Clark Maskell, and Solveig Pederson. Hal Cox noticed that the mathematical properties of targeted mobile genetic elements with genetic payloads derived by the author are identical to Lindenmayer or L-systems and referred the author to Aristid Lindenmayer's research. Mention in these acknowledgements does not indicate endorsement of the paper's suggestions or conclusions.